\documentclass[journal,10pt]{IEEEtran}
\usepackage[T1]{fontenc}
\usepackage{amsmath,bm,amsfonts,amssymb,graphicx,color,mathtools,
setspace,comment,multirow,xfrac}
\usepackage{caption}
\usepackage{subcaption}
\usepackage{booktabs}
\usepackage{epstopdf}
\usepackage{mathtools}
\usepackage{cite}
\usepackage{citesort}
\newcommand*{\QED}{\hfill\ensuremath{\blacksquare}}

\begin{document}
\title{Impact of Line-of-Sight and Unequal Spatial Correlation on 
Uplink MU-MIMO Systems}
\author{Harsh~Tataria,~\IEEEmembership{Member,~IEEE,} 
Peter~J.~Smith,~\IEEEmembership{Fellow,~IEEE,}
Larry~J.~Greenstein,~\IEEEmembership{Life Fellow,~IEEE,}
Pawel~A.~Dmochowski,~\IEEEmembership{Senior~Member,~IEEE,}
and Michail~Matthaiou,~\IEEEmembership{Senior~Member, IEEE}
\vspace{-25pt}
\thanks{H.~Tataria and M.~Matthaiou are with the School of Electronics, Electrical 
Engineering and Computer Science, Queen's University Belfast, Belfast, BT3 9DT, UK (e-mail: \{h.tataria, m.matthaiou\}@qub.ac.uk).}
\thanks{P.~J.~Smith is with the School of Mathematics and Statistics, 
Victoria University of Wellington, Wellington 6140, New Zealand 
(e-mail: peter.smith@vuw.ac.nz).}
\thanks{P.~A.~Dmochowski is with the School of 
Engineering and Computer Science, Victoria University of Wellington, 
Wellington 6140, New Zealand (e-mail: pawel.dmochowski@ecs.vuw.ac.nz).}
\thanks{L.~J.~Greenstein was with the Wireless Information Network
Laboratory, Rutgers University, North Brunswick, NJ 08902, USA (e-mail:
ljg@winlab.rutgers.edu).}}

\maketitle

\begin{abstract}
Closed-form approximations of the expected per-terminal 
signal-to-interference-plus-noise-ratio (SINR) and ergodic sum spectral 
efficiency of a multiuser multiple-input multiple-output system are 
presented. Our analysis assumes spatially correlated Ricean fading channels 
with maximum-ratio combining on the uplink. Unlike previous studies, 
our model accounts for the presence of unequal correlation matrices, 
unequal Rice factors, as well as unequal link gains to each terminal. 
The derived approximations lend themselves to useful insights, 
special cases and demonstrate the aggregate impact of line-of-sight 
(LoS) and unequal correlation matrices. Numerical 
results show that while unequal correlation matrices enhance the 
expected SINR and ergodic sum spectral efficiency, the 
presence of strong LoS has an opposite effect. Our 
approximations are general and remain insensitive to changes in the system 
dimensions, signal-to-noise-ratios, LoS levels and unequal correlation levels. 
\vspace{-3pt}
\end{abstract}

\IEEEpeerreviewmaketitle

\vspace{-15pt}
\section{Introduction}
\label{introduction}
\vspace{-1pt}
The lack of rich scattering and  
insufficient antenna spacing at a cellular base station (BS) leads 
to increased levels of spatial correlation \cite{RUSEK2013}. 
For multiuser multiple-input multiple-output (MU-MIMO) systems, 
this is known to negatively impact the 
signal-to-interference-plus-noise-ratio (SINR) of a given terminal, 
as well as the sum spectral efficiency of the system. 
Numerous works have investigated the SINR and spectral efficiency 
performance of MU-MIMO systems with spatial correlation (see e.g., 
\cite{HOYDIS2013,ZHANG2016,NAM2017} and references 
therein). However, very few of the above mentioned studies consider the 
effects of line-of-sight (LoS) components, likely to be a dominant feature in 
future wireless access with the rise of smaller cell sizes \cite{TATARIA2017}. 
Thus, understanding the performance of such systems with Ricean fading is of 
particular importance. The uplink Ricean analysis presented in 
\cite{ZHANG2014} does not consider the effects of spatial correlation at the BS.  
On the other hand, the related literature (see e.g., 
\cite{FALCONET2016,ZHANG2016}) routinely assumes that on the uplink, 
all terminals are seen by the BS via the same set of incident directions, 
resulting in equal correlation structures. In reality, a different set of 
incident directions are likely to be observed by multiple terminals, due to their 
different geographical locations, leading to variations in the local 
scattering. This gives rise to wide variations in the correlation patterns 
across multiple terminals \cite{NAM2017}. Hence, we consider 
unequal correlation matrices from each terminal.

Motivated by this, with a uniform linear array (ULA) and
maximum-ratio combining (MRC) at the BS, we present
insightful closed-form approximations of the expected perterminal
SINR and ergodic sum spectral efficiency of an uplink
MU-MIMO system. Unlike previous results, for both microwave
and millimeter-wave (mmWave) propagation parameters,
the closed-form expressions consider unequal correlation
matrices, Rice ($K$) factors and link gains for each terminal.
The approximations are shown to be extremely tight for small
and large system dimensions, as well as, arbitrary signal-to-noise-
ratios (SNRs). To the best of our knowledge, this level
of accuracy over such a general channel model capturing a
wide range of scenarios has not been achieved previously.
Numerical results show the aggregate impact of LoS and
unequal spatial correlation. Special cases are presented for
Rayleigh fading channels with equal and unequal correlation
matrices, as well as, for Ricean fading channels with equal
correlation matrices.

\vspace{-11pt}
\section{System Model}
\label{systemmodel}
\vspace{-1pt}
The uplink of a MU-MIMO system operating in an 
urban microcellular environment (UMi) is considered. 
The BS is located at the center of a circular 
cell with radius $R_{c}$, and is equipped with a $M$ element ULA 
simultaneously communicating with 
$L$ single-antenna terminals ($M\gg{}L$). 
Channel knowledge is assumed at the BS, as the prime focus of the manuscript is
on performance analysis with general fading channels and not on system level
imperfections.

The composite $M\times{}1$ received signal at the BS is given by
$\bm{y}=\rho^{\frac{1}{2}}\bm{G}\bm{D}^{\frac{1}{2}}\bm{s}+\bm{n}$, 
where $\rho$ is the average uplink transmit power, $\bm{G}$ is the 
$M\times{}L$ fast-fading channel 
matrix between the $M$ BS antennas and $L$ terminals, $\bm{D}$ is an $L\times{}L$ 
diagonal matrix of link gains, where the link gain for terminal $l$ is 
given by 
$\left[\bm{D}\right]_{l,l}=\beta_{l}$. The large-scale fading effects for 
terminal $l$ in geometric attenuation and shadow-fading 
are captured in 
$\beta_{l}=\varrho{}\zeta_{l}\left(r_{0}/r_{l}\right)^{\alpha}$. In 
particular, $\varrho$ is the unit-less constant for geometric attenuation at 
a reference distance of $r_{0}$, $r_{l}$ is the distance between the 
$l$-th terminal and the BS, $\alpha$ is the attenuation exponent and 
$\zeta_{l}$ captures the effects of shadow-fading, modeled via a log-normal 
density, i.e., $10\log_{10}\left(\zeta_{l}\right)\sim{}\mathcal{N}
\left(0,\sigma_{\textrm{sh}}^{2}\right)$. 
Moreover, $\bm{s}$ is the $L\times{}1$ vector 
of uplink data symbols from $L$ terminals to the BS, such that the 
$l$-th entry of $\bm{s}$, $s_{l}$ has an expected value of one, i.e.,
$\mathbb{E}\left[|s_{l}|^{2}\right]=1$.
The $M\times{}1$ vector of additive white Gaussian noise at the BS is denoted 
by $\bm{n}$, such that the $l$-th entry of 
$\bm{n}$, $n_{l}\sim{}\mathcal{CN}\left(0,\sigma^{2}\right)$. 
We assume that $\sigma^{2}=1$. Hence, the average 
uplink SNR is defined as $\rho/\sigma^{2}=\rho$.  
The $M\times{}1$ channel vector from terminal $l$ to the BS is denoted by 
$\bm{g}_{l}$, which forms the $l$-th column of 
$\bm{G}=\left[\bm{g}_{1},\dots{},\bm{g}_{L}\right]$. 

More specifically, 
\vspace{-9pt}
\begin{equation}
\label{channelmodel}
\bm{g}_{l}=\eta_{l}\bar{\bm{h}}_{l}+\gamma_{l}\bm{R}_{l}^{\frac{1}{2}}
\tilde{\bm{h}}_{l}. 
\vspace{-5pt}
\end{equation}
The $M\times{}1$ LoS and the non LoS (NLoS) 
components of the channel are denoted by 
$\bar{\bm{h}}_{l}$ and $\tilde{\bm{h}}_{l}$. Note that 
$\gamma_{l}=(1/\left(1+K_{l}\right))^{1/2}$ and $\eta_{l}=
(K_{l}/\left(K_{l}+1\right))^{1/2}$, with 
$K_{l}$ being the Ricean $K$-factor for the $l$-th terminal. 
$\bm{R}_{l}$ is the receive correlation matrix specific to terminal $l$, 
$\tilde{\bm{h}}_{l}\sim\mathcal{CN}\left(0,\bm{I}_{M}\right)$ and 
$\bar{\bm{h}}_{l}=[1, e^{j2\pi{}d\cos\left(\phi'_{l}\right)},
\dots{},e^{j2\pi{}d\left(M-1\right)\cos\left(\phi'_{l}\right)}]$. 
Here, $d$ is the equidistant inter-element antenna spacing normalized by 
the carrier wavelength and $\phi'_{l}\sim{}\mathcal{U}
\left[0,2\pi\right]$ is the azimuth angle-of-arrival of the LoS 
component for the $l$-th terminal. 

We employ a linear receiver at the BS array in the form of a MRC filter, 
where $\bm{G}^{\textrm{H}}$ is the $L\times{}M$ filter matrix used to 
separate $\bm{y}$ into $L$ data streams by 
$\bm{r}=\bm{G}^{\textrm{H}}\bm{y}=\rho^{1/2}\bm{G}^{\textrm{H}}
\bm{G}\bm{D}^{1/2}\bm{s}+\bm{G}^{\textrm{H}}\bm{n}$. 
Hence, the combined signal from terminal $l$ is 
given by $r_{l}=\rho^{1/2}\beta_{l}^{1/2}\bm{g}_{l}^{\textrm{H}}\bm{g}_{l}
s_{l}+\rho^{1/2}\sum\nolimits_{\substack{k=1\\k\neq{}l}}^{L}\beta_{k}^{1/2}
\bm{g}_{l}^{\textrm{H}}\bm{g}_{k}s_{k}+\bm{g}_{l}^{\textrm{H}}\bm{n}$. Thus, 
the corresponding SINR for terminal $l$ is given by 
\vspace{-5pt}
\begin{equation}
\label{sinrterminall}
\textrm{SINR}_{l}=\frac{\rho\beta_{l}||\bm{g}_{l}||^{4}}{||\bm{g}_{l}||^{2}+
\rho\sum\nolimits_{\substack{k=1\\k\neq{}l}}^{L}\beta_{k}
|\bm{g}_{l}^{\textrm{H}}\bm{g}_{k}|^{2}}. 
\vspace{-5pt}
\end{equation}
As such, the instantaneous uplink spectral efficiency for the $l$-th terminal 
(measurable in bits/sec/Hz) is given by 
$\textrm{R}_{l}^{\textrm{se}}=\log_{2}\left(1+\textrm{SINR}_{l}\right)$. From here, 
the ergodic sum spectral efficiency over all $L$ terminals is given by 
\vspace{-4pt}
\begin{equation}
\label{ergodicsumspectralefficiency}
\mathbb{E}\left[\textrm{R}^{\textrm{sum}}\right]=
\mathbb{E}\left[\sum\nolimits_{l=1}^{L}
\hspace{-2pt}\textrm{R}_{l}^{\textrm{se}}\right],  
\vspace{-4pt}
\end{equation}
where the expectation is performed over the fast-fading.

\vspace{-9pt}
\section{Expected Per-Terminal SINR and Ergodic Sum Spectral 
Efficiency Analysis}
\label{expectedsinrandergodicspectralefficiencyanalysis}
\vspace{-1pt}
The expected SINR of terminal $l$ can be obtained by evaluating the 
expected value of the ratio in \eqref{sinrterminall}. Exact evaluation 
of this is extremely cumbersome, as shown in \cite{ZHANG2014}. Hence, we 
resort to the first-order Delta method expansion, as shown in the analysis 
methodology of \cite{ZHANG2014}. This gives 
\vspace{-3pt}
\begin{equation}
\label{expectedsinrterminall}
\mathbb{E}\left[\textrm{SINR}_{l}\right]\approx\frac{\rho\beta_{l}
\mathbb{E}\left[||\bm{g}_{l}||^{4}\right]}{\mathbb{E}\left[||\bm{g}_{l}||
\right]^{2}+\rho\sum\nolimits_{\substack{k=1\\k\neq{}l}}^{L}\beta_{k}
\mathbb{E}\left[|\bm{g}_{l}^{\textrm{H}}\bm{g}_{k}|^{2}\right]}. 
\vspace{-8pt}
\end{equation}
\textbf{Remark 1.} The approximation in \eqref{expectedsinrterminall} is of 
the form of $\frac{\mathbb{E}\left[X\right]}{\mathbb{E}\left[Y\right]}$. 
The accuracy of such an approximation relies on $Y$ having a 
small standard deviation relative to its mean.
This can be seen by applying a multivariate Taylor 
series expansion of  $\frac{X}{Y}$ around 
$\frac{\mathbb{E}\left[X\right]}{\mathbb{E}\left[Y\right]}$, 
as shown in the methodology of \cite{ZHANG2014}. Both 
$X$ and $Y$ are well suited to this approximation as $M$ and $L$ 
start to increase. This is evident from the presented numerical results in 
Section~\ref{numericalresults}.

In Lemmas 1, 2 and 3 which follow, we derive the 
expected values in the numerator and denominator of 
\eqref{expectedsinrterminall}.

\textbf{Lemma 1.} For a ULA with $M$ receive antennas at the BS, considering a 
correlated Ricean fading channel, $\bm{g}_{l}$, from the $l$-th terminal to 
the BS 
\begin{align}
\nonumber
\delta_{l}\hspace{-1pt}&=\hspace{-1pt}
\mathbb{E}\left[||\bm{g}_{l}||^{4}\right]\hspace{-1pt}=\hspace{-1pt}
\left(\eta_{l}\right)^{4}\left\{M^{2}+
\textrm{tr}\left[\left(\bm{R}_{l}\right)^{2}\right]\right\}\hspace{-1pt}+
\hspace{-1pt}2M^{2}\left(\eta_{l}\right)^{2}\left(\gamma_{l}\right)^{2}\\ 
\label{lemma1}
&+\hspace{-1pt}
2\left(\gamma_{l}\right)^{2}\left(\eta_{l}\right)^{2}
\left[\bar{\bm{h}}_{l}^{\textrm{H}}\bm{R}_{l}\bar{\bm{h}}_{l}\right]+
\left(\gamma_{l}\right)^{4}M^{2},  
\end{align}
where each parameter is defined after 
\eqref{channelmodel}.

\emph{Proof:} See Appendix~\ref{app1}. \QED

\textbf{Lemma 2.} Under the same conditions as Lemma 1, 
\vspace{-6pt}
\begin{align}
\nonumber
\varphi_{l,k}&\hspace{2pt}=
\mathbb{E}\left[|\bm{g}_{l}^{\textrm{H}}\bm{g}_{k}|^{2}\right]
\hspace{-2pt}=\hspace{-2pt}
\left(\eta_{l}\right)^{2}\hspace{-2pt}\left(\eta_{k}\right)^{2}\hspace{-2pt}
\textrm{tr}\left[\bm{R}_{k}\bm{R}_{l}\right]\\ 
\nonumber
&+\left(
\eta_{l}\right)^{2}\hspace{-2pt}\left(\gamma_{k}\right)^{2}\textrm{tr}\left[
\bar{\bm{h}}_{k}^{\textrm{H}}\bm{R}_{l}
\bar{\bm{h}}_{k}\right]
+\left(\gamma_{l}\right)^{2}\hspace{-2pt}\left(\eta_{k}\right)^{2}
\textrm{tr}\left[
\bar{\bm{h}}_{l}\bar{\bm{h}}_{l}^{\textrm{H}}\bm{R}_{k}
\right]\hspace{-2pt}
\\ \label{lemma2} 
&+\left(\gamma_{l}\right)^{2}\hspace{-2pt}
\left(\gamma_{k}\right)^{2}|\bar{\bm{h}}_{l}^{\textrm{H}}
\bar{\bm{h}}_{k}|^{2}. 
\end{align}

\vspace{-6pt}
\emph{Proof:} See Appendix~\ref{app2}. \QED

\textbf{Lemma 3.} Under the same conditions as Lemma 1, 
\vspace{-5pt}
\begin{equation}
\label{lemma3}
\chi_{l}=
\mathbb{E}\left[||\bm{g}_{l}||^{2}\right]=M\left[\left(\gamma_{l}\right)^{2}+
\left(\eta_{l}\right)^{2}\right]=M. 
\vspace{-7pt}
\end{equation} 

\emph{Proof:} We begin by recognizing that $\chi_{l}=
\mathbb{E}\left[||\bm{g}_{l}||^{2}\right]=\mathbb{E}
\big[\bm{g}_{l}^{\textrm{H}}\bm{g}_{l}\big].$
Substituting the definition of $\bm{g}_{l}$ into \eqref{lemma3} and 
performing the expectations in with respect to 
$\tilde{\bm{h}}$ yields the desired result. Only a 
sketch of the proof is given here, as it relies on straightforward 
algebraic manipulations.\QED

\textbf{Theorem 1.} With MRC and a ULA at the BS, the expected uplink SINR of 
terminal $l$ undergoing spatially correlated Ricean fading can be 
approximated as 
\vspace{-4pt}
\begin{equation}
\label{theorem1}
\mathbb{E}\left[\textrm{SINR}_{l}\right]\approx\frac{\rho\beta_{l}\delta_{l}}
{\chi_{l}+\rho\sum\nolimits_{\substack{k=1\\k\neq{}l}}^{L}\beta_{k}
\varphi_{l,k}},
\vspace{-4pt}
\end{equation}
where $\delta_{l}, \varphi_{l,k}$ and $\chi_{l}$ are given by 
\eqref{lemma1}, \eqref{lemma2} and \eqref{lemma3}, respectively. 

\emph{Proof:} Substituting the results from Lemmas 1, 2 and 3 for 
$\delta_{l}, \chi_{l}$ and $\varphi_{l,k}$ yields 
the desired expression. \QED

\textbf{Remark 2.} Further algebraic manipulations allows us to 
express \eqref{theorem1} as \eqref{theorem1insights}, 
shown on top of the next page for reasons of space. Note that 
\eqref{theorem1insights} can be used to approximate the ergodic 
sum spectral efficiency of the system by stating 
\vspace{-4pt}
\begin{equation}
\label{ergodicsumspectralefficiencyapprox}
\mathbb{E}\left[\textrm{R}^{\textrm{sum}}\right]
\approx\sum\nolimits_{l=1}^{L}\log_{2}\big(1+
\mathbb{E}\left[\textrm{SINR}_{l}\right]\big). 
\vspace{-4pt}
\end{equation}
While the accuracy of \eqref{theorem1insights} and 
\eqref{ergodicsumspectralefficiencyapprox} is demonstrated in 
Section~\ref{numericalresults}, in the sequel, we present the 
implications and special cases of \eqref{theorem1insights} to 
demonstrate its generality.

\vspace{-12pt}
\section{Implications and Special Cases}
\label{implicationsandspecialcases}
\vspace{-3pt}
\subsection{Implications of \eqref{theorem1insights}}
\label{implicationsoftheorem1}
\begin{figure*}[!t]
\vspace{-10pt}
\begin{equation}
\label{theorem1insights}
\mathbb{E}\left[\textrm{SINR}_{l}\right]\approx\frac{\frac{\rho\beta_{l}}
{\left(K_{l}+1\right)^{2}}
\left\{M^{2}\left(1+2K_{l}+K_{l}^{2}\right)+\textrm{tr}\left[\bm{R}_{l}^{2}
\right]+2K_{l}\bar{\bm{h}}_{l}^{\textrm{H}}
\bm{R}_{l}\bar{\bm{h}}_{l}\right\}}{M+\rho\sum\nolimits_{k=1, k\neq{}l}^{L}
\frac{\beta_{k}}{\left(
K_{k}+1\right)\left(K_{l}+1\right)}\left\{\textrm{tr}\left[\bm{R}_{k}
\bm{R}_{l}\right]+
K_{k}\left(\bar{\bm{h}}_{k}^{\textrm{H}}\bm{R}_{l}\bar{\bm{h}}_{k}\right)
+K_{l}\left(
\bar{\bm{h}}_{l}^{\textrm{H}}\bm{R}_{k}\bar{\bm{h}}_{l}\right)+K_{l}K_{k}
\left|\bar{\bm{h}}_{l}^{\textrm{H}}\bar{\bm{h}}_{k}\right|^{2}\right\}}. 
\vspace{-5pt}
\end{equation}
\hrulefill
\vspace{-15pt}
\end{figure*}

\vspace{-3pt}
Both the numerator and the denominator of \eqref{theorem1insights} 
contain quadratic forms of the type $\bar{\bm{h}}^{\textrm{H}}
\bm{R}\bar{\bm{h}}$. Via the Rayleigh quotient result, such 
quadratic forms are maximized when $\bar{\bm{h}}$ is parallel (aligned) 
to the maximum eigenvector of $\bm{R}$. From this, an interesting 
observation can be made: Alignment of $\bar{\bm{h}}_{l}$ and 
$\bm{R}_{l}$ amplifies the expected signal power, while alignment of 
$\bar{\bm{h}}_{k}$ with $\bm{R}_{l}$, $\bar{\bm{h}}_{l}$ with $\bm{R}_{k}$ 
and $\bar{\bm{h}}_{l}$ with $\bar{\bm{h}}_{k}$ increases the expected 
interference power, leading to a lower SINR. Likewise, if 
$\bm{R}_{k}$ and $\bm{R}_{l}$ become similar, then 
$\textrm{tr}\left[\bm{R}_{k}\bm{R}_{l}\right]$ increases, degrading 
the SINR. The global observation is that the SINR 
reduces by virtue of channel similarities of various types (LoS 
and correlation) and increases if the channels are more diverse.

\vspace{-14pt}
\subsection{Special Cases of \eqref{theorem1insights}}
\label{specialcasesoftheorem1}
\vspace{-3pt}
\textbf{Corollary 1.} In pure NLoS conditions (i.e., Rayleigh 
fading) with unequal correlation matrices, 
\eqref{theorem1insights} reduces to  
\begin{equation}
\label{corollary1}
\mathbb{E}\left[\textrm{SINR}_{l}^{\textrm{c1}}\right]\approx
\frac{\rho\beta_{l}\left\{M^{2}+\textrm{tr}\left[\bm{R}_{l}^{2}\right]
\right\}}{M+\rho\sum\nolimits_{\substack{k=1\\k\neq{}l}}^{L}\beta_{k}
\Big\{\textrm{tr}\left[\bm{R}_{k}
\hspace{-1pt}\bm{R}_{l}\right]\Big\}}. 
\end{equation}

\emph{Proof:} Substituting $K_{l}=K_{k}=0, 
\forall{}l,k=\left\{1,\dots{},L\right\}$ in \eqref{theorem1insights} yields 
the desired result. \QED

\textbf{Corollary 2 (Proposition 1 in \cite{ZHANG2016}).} 
In pure Rayleigh fading with equal correlation 
matrices, \eqref{theorem1insights} collapses to
\begin{equation}
\label{corollary2}
\mathbb{E}\left[\textrm{SINR}_{l}^{\textrm{c2}}\right]\approx
\frac{\rho\beta_{l}\left\{M^{2}+\textrm{tr}\left[\bm{R}_{l}^{2}\right]
\right\}}{M+\rho\sum\nolimits_{\substack{k=1\\k\neq{}l}}^{L}\beta_{k}
\Big\{\textrm{tr}\left[\bm{R}_{l}^{2}
\right]\Big\}}.
\end{equation}

\emph{Proof:} Setting $\bm{R}_{l}=\bm{R}_{k},\forall{}l,k=\left\{
1,\dots{},L\right\}$ in \eqref{corollary1} gives the desired result. The 
result is consistent with \cite{ZHANG2016}. \QED

\textbf{Corollary 3.} With LoS presence and equal correlation matrices, 
\eqref{theorem1insights} can be approximated with
\vspace{-5pt} 
\begin{equation}
\label{corollary3} 
\vspace{-6pt}
\mathbb{E}\left[\textrm{SINR}_{l}^{\textrm{c3}}\right]\approx
\frac{\rho\beta_{l}\delta_{l}}{\chi_{l}+
\rho\sum\nolimits_{\substack{k=1\\k\neq{}l}}^{L}
\beta_{k}\widetilde{\varphi}_{l,k}}, 
\end{equation}
where 
$\widetilde{\varphi}_{l,k}\hspace{-2pt}=\hspace{-2pt}
\left(\eta_{l}\right)^{2}\hspace{-2pt}\left(\eta_{k}\right)^{2}\hspace{-2pt}
\textrm{tr}\left[\bm{R}_{l}^{2}\right]\hspace{-1pt}+\hspace{-2pt}\left(
\eta_{l}\right)^{2}\hspace{-2pt}\left(\gamma_{k}\right)^{2}\hspace{-1pt}
\textrm{tr}\left[
\bar{\bm{h}}_{k}^{\textrm{H}}\hspace{-1pt}\bm{R}_{l}\hspace{-1pt}
\bar{\bm{h}}_{k}\right]\hspace{-1pt}+\hspace{-1pt}
\left(\gamma_{l}\right)^{2}\hspace{-2pt}\left(\eta_{k}\right)^{2}
\hspace{-1pt}\textrm{tr}\left[
\bar{\bm{h}}_{l}\bar{\bm{h}}_{l}^{\textrm{H}}\hspace{-1pt}\bm{R}_{l}
\right]\hspace{-2pt}
+\hspace{-1pt}\left(\gamma_{l}\right)^{2}\hspace{-2pt}
\left(\gamma_{k}\right)^{2}\hspace{-1pt}|\bar{\bm{h}}_{l}^{\textrm{H}}
\bar{\bm{h}}_{k}|^{2}$.

\emph{Proof:} Replacing $\bm{R}_{k}$ with $\bm{R}_{l}$ and substituting the 
definition of $\delta_{l}$ and $\chi_{l}$ from  \eqref{lemma1} and 
\eqref{lemma3} yields the desired result. \QED

\vspace{-13pt}
\section{Numerical Results}
\label{numericalresults}
\vspace{-2pt}
We employ a statistical approach to determine whether a given terminal 
experiences LoS or NLoS propagation. The NLoS and LoS probabilities are 
governed by the link distance, from which other link parameters such as the 
attenuation exponent and shadow-fading standard deviation are selected. We 
consider the UMi propagation parameters for microwave \cite{3GPPTR36873} and 
mmWave \cite{AKDNENIZ2014,THOMAS2014} frequencies at 2 and 28 GHz, 
respectively. For both cases, the cell radius 
$\left(R_{c}\right)$ and exclusion area $\left(r_{0}\right)$ are fixed to 100 
m and 10 m. The terminals are randomly located outside $r_{0}$ and inside $R_{c}$ 
with a uniform distribution with respect to the cell area. The LoS and NLoS attenuation 
exponents $\left(\alpha\right)$ are given by 2.2, 3.67 and 2, 2.92 at 
microwave and mmWave frequencies, while the parameter 
$\varrho_{l}$ is chosen such that the fifth percentile of the instantaneous 
SINR of terminal $l$ is $0$ dB at $\rho=0$ dB, for the system dimensions of 
$M=64, L=4$. Moreover, the LoS and NLoS shadow-fading standard deviations 
$\left(\sigma_{\textrm{sh}}\right)$ are 3 dB, 4 dB and 5.8 dB, 8.7 dB for the 
microwave and mmWave cases. 
The Ricean $K$-factor has a log-normal density 
with a mean of $9$ and standard deviation of $5$ dB for microwave 
$\left(K_{l}\sim{}\textrm{ln}\left(9,5\right)\right)$ 
\cite{3GPPTR36873} and a mean of $12$ with standard deviation of $3$ dB for 
the mmWave $\left(K_{l}\sim{}\textrm{ln}\left(12,3\right)\right)$ cases 
\cite{THOMAS2014}. With microwave parameters, the probability of terminal $l$ 
experiencing LoS is given by 
$P_{\textrm{LoS}}\left(r_{l}\right)=(\min(18/r_{l},1)
(1-e^{-r_{l}/36}))+e^{-r_{l}/36}$\cite{3GPPTR36873}. Equivalently, at mmWave, 
$P_{\textrm{LoS}}=(1-P_{\textrm{out}\left(r_{l}\right)})
e^{-\iota_{\textrm{LoS}}r_{l}}$, where $1/\iota_{\textrm{LoS}}=67.1$ m and 
$P_{\textrm{out}}$,
the outage probability, is set to $0$ for simplicity \cite{AKDNENIZ2014}.
For both cases, $P_{\textrm{NLoS}}=1-P_{\textrm{LoS}}$. 
Due to its generality in modeling spatially correlated fading, 
the one-ring model is chosen to generate unequal spatial correlation at the BS, as in 
\cite{HOYDIS2013,NAM2017,JIANG2015}.
The $\left(i,j\right)$ entry in the 
correlation matrix of terminal $l$ is given by \cite{JIANG2015}
\vspace{-6pt}
\begin{equation}
\label{oneringcorr}
\left[\bm{R}_{l}\right]_{i,j}=\frac{1}{2\Delta}
\int\nolimits_{-\Delta+\phi_{l}}^{\Delta+\phi_{l}}e^{-j2\pi{}d\left(i-j\right)
\sin\left(\theta_{l}\right)}d\theta_{l}, 
\vspace{-5pt}
\end{equation}
where $\Delta$ denotes the azimuth angular spread, $\phi_{l}$ is the central 
azimuth angle from terminal $l$ to the BS array, $\theta_{l}$ is the actual 
angle-of-arrival (AoA) and $d\left(i-j\right)$ captures the inter-element 
spacing normalized by the carrier wavelength between $i$-th and $j$-th 
antenna elements. Unless explicitly stated, 
we set $d\left(1\right)=0.5$ and assume that $\phi_{l}\sim{}\mathcal{U}
\left[0,2\pi\right]$. The instantaneous value of $\theta_{l}$ is also drawn from a 
uniform distribution on $\frac{-\Delta}{2},\frac{\Delta}{2}$, i.e., 
$\theta_{l}\sim\mathcal{U}\left[\frac{-\Delta}{2},\frac{\Delta}{2}\right]$. 
As such, $\Delta$ represents the total angular spread, naturally bounded from 
$0$ to $2\pi$ radians $(0$ to $360^{\circ})$.  
Note that the one-ring model captures a general 
physical scenario and is not intended to be specific for a particular carrier 
frequency. Naturally, one can fix $d\left(1\right)$ and the distribution of 
$\phi_{l}$, and select values for $\Delta$ from channel measurements at both 
microwave and mmWave frequencies. However, $\Delta$ is varied delibrately to 
understand its impact with LoS on the expected SINR and ergodic sum spectral 
efficiency.

\begin{figure}[!t]
\vspace{-17pt}
 \centering
 \includegraphics[width=7.8cm]{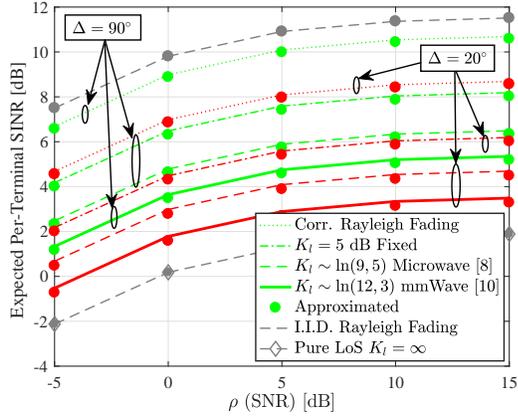}
 \vspace{-5pt}
\caption{Expected per-terminal SNR vs. $\rho$ (SNR) with $M=32, L=3$ and 
 $\Delta=20^{\circ}$ and $90^{\circ}$.}
 \label{expSINRvsSNR}
\vspace{-11pt}
\end{figure}

With $M=32, L=3$, Fig.~\ref{expSINRvsSNR} illustrates the expected 
per-terminal SINR of a given terminal as a function of $\rho$. In addition to 
the microwave and mmWave cases, we consider the two extremes in  
uncorrelated Rayleigh fading and pure LoS channels. Furthermore, 
unequally correlated Rayleigh and Ricean fading cases are considered, where 
the Ricean case has a fixed $K$-factor of $5$ dB for each terminal. 
Three trends can be observed: (1) Transitioning from 
larger to smaller angular spread ($\Delta=90^{\circ}$ to $\Delta=20^{\circ}$) 
significantly reduces the expected SINR for all cases. This is 
despite the fact that the ULA is equipped 
with a moderate number of receive antennas, and is due to the reduction in 
the spatial selectivity of the channel, enforcing the ULA to see a narrower 
spread of the incoming power. (2) Increasing the mean of $K$ has a negative 
impact on the expected SINR, as stronger LoS presence tends to reduce the 
multipath diversity and the rank of the composite channel. 
(3) The proposed expected SINR approximations in 
\eqref{theorem1insights} are seen to remain extremely tight for the entire 
range of $\rho$ for all cases. The approximations can also be seen to 
remain tight for the special case of Rayleigh fading with unequal 
correlation matrices in \eqref{corollary1}. Furthermore, the expected 
SINR in each case is seen to saturate with $\rho$, as the MRC filter is 
unable to mitigate multiuser interference. 

\begin{figure}[!t]
\vspace{-5pt}
 \centering
 \includegraphics[width=7.9cm]{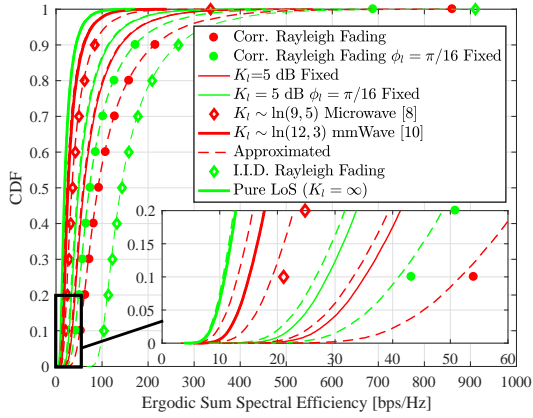}
 \vspace{-5pt}
 \caption{Ergodic sum spectral efficiency CDF with $M=256, L=32$ at 
 $\rho=10$ dB and $\Delta=20^{\circ}$.}
 \label{ergSSECDF}
\vspace{-18pt}
\end{figure}

Considering the special cases in \eqref{corollary2} and \eqref{corollary3}, 
we now examine the aggregate impact of LoS, as well as 
equal and unequal correlation on the ergodic sum spectral 
efficiency, as shown in Fig.~\ref{ergSSECDF}. With $M=256$ and $L=32$, 
using the same propagation parameters as in Fig.~\ref{expSINRvsSNR}, at 
$\rho=10$ dB, we compare the cumulative distribution functions (CDFs) of 
the derived ergodic sum spectral efficiency approximation in 
\eqref{ergodicsumspectralefficiencyapprox} with its simulated counterparts. 
Each CDF is obtained by averaging over the fast-fading, with each value 
representing the variations in the link gains and the $K$-factors. 
The derived approximations remain tight with changes in the 
system size.
Moreover, irrespective of the underlying propagation characteristics, 
unequal correlation matrices result in higher ergodic 
sum spectral efficiency, allowing the ULA to leverage more spatial 
diversity. This is noticed when comparing the $K_{l}=5$ dB curves 
with a fixed $\phi_{l}=\pi/16$ (equal correlation) and variable 
$\phi_{l}$ (unequal correlation) for each terminal. In contrast to the 
correlated Rayleigh case, a dominant LoS component is again seen to be 
detrimental to system performance. 

\vspace{-16pt}
\section{Conclusion}
\label{conclusion}
\vspace{-6pt}
We have presented a general, yet insightful approximation to the 
expected per-terminal SINR and ergodic sum spectral efficiency of an 
uplink MU-MIMO system. With a ULA and MRC at the BS, the approximation is 
robust to equal and unequal correlation matrices, unequal levels of LoS, 
unequal link gains, unequal operating SNRs and system dimensions. With 
both microwave and mmWave parameters, our results show that unequal 
correlation matrices yield higher expected SINRs and ergodic sum spectral 
efficiency in comparison to equal correlation. Moreover, increasing the 
LoS component of the channel reduces the expected SINR and ergodic sum 
spectral efficiency due to the loss of spatial diversity.

\vspace{-15pt}
\appendices
\section{Proof of Lemma 1}
\label{app1}
\vspace{-3pt}
We begin by recognizing that $\delta_{l}=
\mathbb{E}\left[||\bm{g}_{l}||^{4}\right]=
\mathbb{E}[\left(||\bm{g}_{l}||^{2}\right)^{2}]$. Substituting the 
definition of $\bm{g}_{l}$ and denoting $\bm{v}_{l}=\gamma_{l}
\bm{R}_{l}^{\frac{1}{2}}\tilde{\bm{h}}_{l}$ 
and $\bm{q}_{l}=\eta_{l}\bar{\bm{h}}_{l}$ allows us to state 
\vspace{-4pt}
\begin{equation}
\label{lemma1proof1}
\delta_{l}\hspace{-1pt}=\hspace{-1pt}
\mathbb{E}\left[\left(||\bm{g}_{l}||^{2}\right)^{2}\right]
\hspace{-2pt}=\hspace{-2pt}\mathbb{E}
\left[\left(\bm{v}_{l}^{\textrm{H}}
\bm{v}_{l}\hspace{-1pt}+\hspace{-1pt}
\bm{v}_{l}^{\textrm{H}}\bm{q}_{l}\hspace{-1pt}+\hspace{-1pt}
\bm{q}_{l}^{\textrm{H}}\bm{v}_{l}\hspace{-1pt}+\hspace{-1pt}
\bm{q}_{l}^{\textrm{H}}\bm{q}_{l}\right)^{2}\right]\hspace{-2pt}. 
\vspace{-3pt}
\end{equation}
Expanding \eqref{lemma1proof1} allows us to write 
\begin{align}
\nonumber
\delta_{l}&=\hspace{1pt}\mathbb{E}\left[
\left(\bm{v}_{l}^{\textrm{H}}\bm{v}_{l}\right)^{2}\hspace{-2pt}
+\hspace{-2pt}2\left(\bm{v}_{l}^{\textrm{H}}
\bm{v}_{l}\right)\left(\bm{q}_{l}^{\textrm{H}}\bm{q}_{l}\right)\hspace{-2pt}
+\hspace{-2pt}
\left(\bm{v}_{l}^{\textrm{H}}\bm{q}_{l}\bm{q}_{l}^{\textrm{H}}\bm{v}_{l}
\right)\hspace{-1pt}\right.\\ 
\label{lemma1proof2}
&\left.+\left(\bm{q}_{l}^{\textrm{H}}\bm{v}_{l}\bm{v}_{l}^{\textrm{H}}
\bm{q}_{l}\right)+
\left(\bm{q}_{l}^{\textrm{H}}\bm{q}_{l}\right)^{2}\right]. 
\end{align}
Performing the expectations over $\bm{v}_{l}$ in the last four terms of 
\eqref{lemma1proof2} and simplifying yields 
\vspace{-2pt}
\begin{align}
\nonumber
\delta_{l}&=
\mathbb{E}\left[\left(\bm{v}_{l}^{\textrm{H}}\bm{v}_{l}\right)^{2}\right]+
2M\left(\eta_{l}\right)^{2}\hspace{-2pt}\left(\bm{q}_{l}^{\textrm{H}}
\bm{q}_{l}\right)\\ 
\label{lemma1proof3}
&+2\left(\eta_{l}\right)^{2}\bm{q}_{l}^{\textrm{H}}
\bm{R}_{l}\bm{q}_{l}+\left(\bm{q}_{l}^{\textrm{H}}\bm{q}_{l}\right)^{2}. 
\end{align}
After noting that $\mathbb{E}\big[\big(\bm{v}_{l}^{\textrm{H}}\bm{v}_{l}
\big)^{2}\big]=
\mathbb{E}\left[\bm{v}_{l}^{\textrm{H}}\bm{v}_{l}\bm{v}_{l}^{\textrm{H}}
\bm{v}_{l}\right]$, 
substituting the definition of $\bm{v}_{l}$ and extracting the relevant 
constants yields 
$\mathbb{E}\big[\big(\bm{v}_{l}^{\textrm{H}}\bm{v}_{l}\big)^{2}\big]=
\big(\eta_{l}\big)^{4}\mathbb{E}\big[\big(\tilde{\bm{h}}_{l}^{\textrm{H}}
\bm{R}_{l}\tilde{\bm{h}}_{l}\big)^{2}\big]$, 
where $\bm{R}_{l}=\bm{\Phi\Lambda\Phi}^{\textrm{H}}$ via an eigenvalue 
decomposition. Hence, 
\vspace{-8pt}
\begin{equation}
\label{lemma1proof5}
\mathbb{E}\left[\left(\bm{v}_{l}^{\textrm{H}}\bm{v}_{l}\right)^{2}\right]=
\left(\eta_{l}\right)^{4}\mathbb{E}\left[\hspace{2pt}
\left(\sum\nolimits_{i=1}^{M}\left[\bm{\Lambda}\right]_{i,i}
\left|\left(\tilde{\bm{h}}_{l}\right)_{i}\right|^{2}\right)^{2}\right]. 
\vspace{-10pt}
\end{equation}
Performing the expectation with respect to $\tilde{\bm{h}}_{l}$ and 
simplifying yields $\mathbb{E}[(\bm{v}_{l}^{\textrm{H}}
\bm{v}_{l})^{2}]=(\eta_{l})^{4}\{(\textrm{tr}\left[\bm{R}_{l}])^{2}
+\textrm{tr}\left[\bm{R}_{l}^{2}\right]\right\}$. As 
$\textrm{tr}\left[\bm{R}_{l}\right]=M$, 
$\mathbb{E}[(\bm{v}_{l}^{\textrm{H}}\bm{v}_{l})^{2}]=
\left(\eta_{l}\right)^{4}\{M^{2}+\textrm{tr}[(
\bm{R}_{l})^{2}]\}$. 
Substituting the right-hand side along with the 
definition of $\bm{q}_{l}$ into \eqref{lemma1proof3}, recognizing  
$\bar{\bm{h}}_{l}^{\textrm{H}}\bar{\bm{h}}_{l}=M$ and 
simplifying yields Lemma 1.

\vspace{-6pt}
\section{Proof of Lemma 2}
\label{app2}
\vspace{-3pt}
Applying the definition of $\bm{g}_{l}$ and $\bm{g}_{k}$ 
into $\varphi_{l,k}=\mathbb{E}\left[|
\bm{g}_{l}^{\textrm{H}}\bm{g}_{k}|^{2}\right]$ and denoting 
$\bm{v}_{l}=\eta_{l}\bm{R}_{l}^{\frac{1}{2}}
\tilde{\bm{h}}_{l}$ and $\bm{q}_{l}=
\gamma_{l}
\bar{\bm{h}}_{l}$ yields 
$\varphi_{l,k}=\mathbb{E}\left[\hspace{1pt}\left|
\hspace{1pt}\left(\bm{v}_{l}^{\textrm{H}}+\bm{q}_{l}^{\textrm{H}}\right)
\left(\bm{v}_{k}+\bar{\bm{h}}_{k}
\right)\right|^{2}\right]$. 
Expanding and simplifying further gives  
\vspace{-6pt}
\begin{align}
\nonumber
\varphi_{l,k}&=\hspace{2pt}
\mathbb{E}\left[\bm{v}_{l}^{\textrm{H}}\bm{v}_{k}\bm{v}_{k}^{\textrm{H}}
\bm{v}_{l}\right]+
\mathbb{E}\left[\bm{v}_{l}^{\textrm{H}}\bm{q}_{k}\bm{q}_{k}^{\textrm{H}}
\bm{v}_{l}\right]\\
\label{lemma2proof3}
&+\mathbb{E}\left[\bm{q}_{l}^{\textrm{H}}\bm{v}_{k}\bm{v}_{k}^{\textrm{H}}
\bm{q}_{l}\right]+
\mathbb{E}\left[\bm{q}_{l}^{\textrm{H}}\bm{q}_{k}\bm{q}_{k}^{\textrm{H}}
\bm{q}_{l}\right]. 
\end{align}
Recognizing that $\mathbb{E}\left[\bm{v}_{l}\bm{v}_{l}^{\textrm{H}}\right]=
\mathbb{E}\left[\eta_{l}\bm{R}_{l}^{\frac{1}{2}}
\tilde{\bm{h}}_{l}\eta_{l}\tilde{\bm{h}}_{l}^{\textrm{H}}
\bm{R}_{l}^{\frac{1}{2}}
\right]=\left(\eta_{l}\right)^{2}\hspace{-2pt}\bm{R}_{l}$, substituting 
back the definitions of $\bm{v}_{l}$, $\bm{v}_{k}$, 
$\bm{q}_{l}$ and $\bm{q}_{k}$ in \eqref{lemma2proof3} and extracting the 
relevant constants yields 
\vspace{-8pt}
\begin{align}
\nonumber
\varphi_{l,k}&\hspace{2pt}=
\left(\eta_{l}\right)^{2}\left(\eta_{k}\right)^{2}
\textrm{tr}\left[\bm{R}_{k}\bm{R}_{l}\right]
\\ \nonumber
&+\left(\eta_{l}\right)^{2}\left(\gamma_{k}\right)^{2}
\mathbb{E}\left[\textrm{tr}\left[\bm{R}_{l}^{\frac{1}{2}}
\bar{\bm{h}}_{k}\bar{\bm{h}}_{k}^{\textrm{H}}\bm{R}_{l}^{\frac{1}{2}}
\tilde{\bm{h}}_{l}\tilde{\bm{h}}_{l}^{\textrm{H}}\right]\right]\\ 
\nonumber
&+\left(\gamma_{l}\right)^{2}\left(\eta_{k}\right)^{2}
\mathbb{E}\left[\textrm{tr}\left[\tilde{\bm{h}}_{k}
\tilde{\bm{h}}_{k}^{\textrm{H}}\bm{R}_{k}^{\frac{1}{2}}\bar{\bm{h}}_{l}
\bar{\bm{h}}_{l}^{\textrm{H}}
\bm{R}_{k}^{\frac{1}{2}}\right]\right]\\
\label{lemma2proof4}
&+\left(\gamma_{l}\right)^{2}\left(\gamma_{k}\right)^{2}
|\bar{\bm{h}}_{l}^{\textrm{H}}\bar{\bm{h}}_{k}|^{2}. 
\vspace{-5pt}
\end{align}
Taking the trace and simplifying yields \eqref{lemma2}.

\vspace{-7pt}
\bibliographystyle{IEEEtran}
{}

\end{document}